\RequirePackage{lineno}
\documentclass[twocolumn,letterpaper,aps,prc,longbibliography,superscriptaddress,nofootinbib,floatfix]{revtex4-1}

\usepackage{amsmath}	
\usepackage{graphicx}	
\usepackage{hyperref}
\hypersetup{colorlinks=true, urlcolor=cyan, linkcolor=blue, citecolor=blue}

\usepackage{xspace}	

\begin{document}

\title{Investigation of hadronic effects on resonance productions \\in small collision systems using the EPOS4 model}
%

\newcommand{\pusan}{Department of Physics, Pusan National University, Busan 46241, South Korea}
\affiliation{\pusan}

\newcommand{\yonsei}{Department of Physics, Yonsei University, Seoul 03722, South Korea}
\affiliation{\yonsei}

\newcommand{\tokyo}{Center for Nuclear Study, The University of Tokyo, Tokyo 113-0033, Japan}
\affiliation{\tokyo}


\newcommand{\cern}{European Organization for Nuclear Research (CERN), Geneva 01631, Switzerland}
\affiliation{\cern}

\author{Hyunji Lim} \affiliation{\pusan}
\author{Bong-Hwi Lim} \affiliation{\tokyo}
\author{Minjung Kim} \affiliation{\cern}
\author{Sanghoon Lim} \email{shlim@yonsei.ac.kr} \affiliation{\yonsei}

\date{\today}

\begin{abstract}


Recent experimental results in high-multiplicity proton-proton (pp) collisions have suggested the possible emergence of collective behavior and medium-like effects previously considered characteristic of heavy-ion collisions. Resonance production provides a sensitive probe of such effects, as resonance yields and transverse-momentum distributions can be modified by hadronic interactions occurring between chemical and kinetic freeze-out. In this study, these effects are investigated using the EPOS4 event generator, in which hadronic final-state interactions are modeled through the UrQMD transport approach. By comparing calculations performed with and without UrQMD, the impact of hadronic interactions on resonance production is evaluated. In addition, the UrQMD contributions are separated into regeneration and rescattering, enabling a detailed investigation of both resonance production enhancement and the loss of reconstructible resonance signals. The analysis is performed for various mesonic and baryonic resonances with different lifetimes in pp collisions at LHC energies and is extended to p-O, O-O, and Pb-Pb collisions to study the system-size dependence of hadronic-phase effects. The results show that resonance production is governed by the competition between regeneration and rescattering, whose relative importance depends strongly on the resonance species, transverse momentum, and collision system. While rescattering suppresses reconstructible short-lived resonance signals, regeneration can significantly enhance the yields of several resonance species, particularly baryonic resonances. These findings demonstrate that hadronic interactions can play an important role even in small collision systems and highlight the need to measure resonances with different lifetimes and quantum numbers to constrain the dynamics and lifetime of the hadronic phase across collision systems.

\end{abstract}


\maketitle

\section{Introduction}
\label{sec:Intro}

At high energies, such as those achieved at the Relativistic Heavy Ion Collider (RHIC) and the Large Hadron Collider (LHC), a transient quark–gluon plasma (QGP) is expected to form~\cite{PHENIX:2004vcz, STAR:2005gfr, PHOBOS:2004zne, BRAHMS:2004adc, ALICE:2022wpn}. As this hot and dense medium expands and cools, it hadronizes at a critical temperature of approximately 155–160 MeV, producing stable hadrons and resonances. These particles subsequently populate a hadron gas, where strong interactions continue until the system reaches kinetic freeze-out. This evolution takes place over a timescale of several tens of fm/c, comparable to the lifetimes of many resonances listed in Table~\ref{tab:table_res}. Consequently, some resonances decay within the hadronic phase. Their decay products may rescatter with surrounding hadrons, modifying their momenta and reducing the probability of reconstructing the original resonance. Conversely, pseudo-elastic scattering among hadrons in the hadronic medium can regenerate resonances, partially counteracting the signal suppression caused by rescattering~\cite{Steinheimer:2012era, Steinheimer:2015msa, Steinheimer:2017vju}. The interplay between these two competing mechanisms, namely rescattering-induced signal loss and regeneration-driven yield enhancement, determines the final measurable resonance yield. This balance is highly sensitive to the scattering cross sections of the decay products, the density of the hadronic medium, the resonance lifetime, and the duration of the hadronic phase~\cite{ALICE:2023edr}. Since rescattering effects are expected to be more pronounced in larger systems, where the hadron gas phase lasts longer, resonances provide a powerful probe of the properties of hadronic matter formed in high-energy nuclear collisions~\cite{ALICE:2019xyr, ALICE:2023edr}.


\begin{table*}[tbh]
\begin{ruledtabular}
\caption{\label{tab:table_res}
    Information about the species and characteristics of resonances~\cite{ParticleDataGroup:2024cfk}.}
\begin{tabular}{ccccc}
Resonance               & Quark Content                  & Decay Channel   & B.R   & Lifetime (fm/$c$) \\ \hline
$\rho(770)^0$            & $(u\bar{u}/d\bar{d})/\sqrt{2}$ & $\pi^++\pi^-$   & 1.0   & 1.335             \\
$\mathrm{K}^*(892)^0$             & $d\bar{s}$                     & $\pi^++K^-$     & 0.67  & 4.16              \\
$\phi(1020)$              & $s\bar{s}$                     & $K^++K^-$       & 0.492 & 46.26             \\
$\Delta(1232)^{++}$  & $uuu$                          & $\pi^++p$       & 1.0   & 1.69              \\
$\Sigma(1385)^+$   & $uus$                          & $\pi^++\Lambda$ & 0.870 & 5.48              \\
$\Sigma(1385)^-$    & $dds$                          & $\pi^-+\Lambda$ & 0.870 & 5.01              \\
$\Lambda(1520)$      & $uds$                          & $K^-+p$         & 0.225 & 12.54             \\
$\Xi(1530)^0$      & $uss$                          & $\pi^++\Xi^-$   & 0.67  & 22                
\end{tabular}%
\end{ruledtabular}
\end{table*}

A particularly illuminating example is the system-size dependence of hadronic rescattering established by recent ALICE measurements across pp, p--Pb, Xe--Xe, and Pb--Pb collision systems~\cite{ALICE:2023edr}. These studies show that the rescattering effect on $\mathrm{K}^*(892)^0$ yields is not determined solely by the size of the colliding nuclei, but is primarily driven by the produced charged-particle multiplicity, which serves as a proxy for the volume of matter at chemical freeze-out. This finding supports a unified, multiplicity-driven picture of hadronic-phase dynamics that connects small and large collision systems and motivates extending such investigations to previously unexplored system sizes.

Recently, interest in resonance production has extended to smaller collision systems, such as proton–proton (pp) collisions, motivated by evidence that several phenomena once considered unique to heavy-ion collisions also appear in smaller systems. For example, studies of $\mathrm{K}^{*}(892)^0$ and $\phi(1020)$ production in pp collisions at 13 TeV and p--Pb collisions at 5.02 TeV show that the $\mathrm{K}^{*}(892)^0$-to-kaon yield ratio decreases with increasing charged-particle multiplicity~\cite{ALICE:2019etb, ALICE:2016sak}, following a trend similar to that observed in Xe--Xe and Pb--Pb collisions~\cite{ALICE:2023edr}. This behavior can be attributed to the short lifetime of the $\mathrm{K}^{*}(892)^0$, about 4.2 fm/$c$, which makes it more susceptible to rescattering. In contrast, the $\phi(1020)$, with a much longer lifetime of about 46.3 fm/$c$, shows no comparable multiplicity dependence. Beyond these canonical resonances, recent measurements in pp collisions at 13 TeV demonstrate that other short-lived resonances, including $\rho(770)^0$, $\Sigma(1385)^{\pm}$, and $f_0(980)$, also exhibit multiplicity-dependent yield modifications consistent with hadronic-phase effects~\cite{ALICE:2023egx, ALICE:2025lpf, ALICE:2018qdv}, whereas longer-lived species such as $\Xi(1530)^0$ remain largely unaffected. Taken together, these findings indicate that the resonance lifetime is a key discriminator of hadronic-phase sensitivity across collision systems.

These observations suggest that a hadronic phase can influence resonance production even in small collision systems. The EPOS3 event generator, incorporating UrQMD to describe interactions during the hadronic phase, can reproduce the multiplicity dependence of the $\mathrm{K}^{*}(892)^0/\mathrm{K}$ and $\phi(1020)/\mathrm{K}$ ratios in Pb--Pb collisions~\cite{Knospe:2015nva}. However, a systematic study of such effects in small collision systems, including pp and p--Pb collisions, remains limited.
More recently, EPOS4 has been used to investigate the dynamics of various resonance species in pp and Pb--Pb collisions~\cite{Sumberia:2024oea}, providing further insight into the role of hadronic interactions in modifying resonance production.

This paper investigates resonance production in pp collisions using the EPOS4 model~\cite{Werner:2023zvo}, with particular emphasis on the role of hadronic-phase interactions. EPOS4 represents a significant advancement over its predecessor. 
It employs a dynamical core–corona separation framework, in which high-density regions evolve collectively via hydrodynamics, while low-density regions hadronize via string fragmentation. A microcanonical approach to core hadronization enforces exact quantum-number conservation on an event-by-event basis, providing a more realistic description of particle yields and strangeness enhancement in small systems~\cite{Werner:2023zvo}. 
Importantly, EPOS4 allows the UrQMD hadronic afterburner to be switched on or off, enabling a clean, model-internal separation of hadronic-phase effects from primary particle production. This feature has been exploited in recent EPOS4 studies of resonance and identified-particle production in pp and Pb--Pb collisions, where comparisons between the UrQMD-on and UrQMD-off configurations were used to quantify rescattering, regeneration, baryon–antibaryon annihilation, radial-flow effects, and the multiplicity dependence of the hadronic-phase lifetime~\cite{Knospe:2015nva, Sumberia:2024oea}. Complementary studies based on PYTHIA8 have also investigated rescattering effects in pp and p--Pb collisions through resonance-to-stable-particle yield ratios and double ratios between rescattering-on and rescattering-off configurations, further motivating a systematic model-based study of hadronic interactions in small collision systems~\cite{JI2026123455}.

We compare the modeled production of $\mathrm{K}^{*}(892)^0$ and $\phi(1020)$ resonances with experimental data from the ALICE Collaboration~\cite{ALICE:2019etb}. In addition, a comprehensive study is conducted for the resonance species listed in Table~\ref{tab:table_res}, spanning a wide range of lifetimes and quark contents, to systematically explore how hadronic interactions affect different particles. In high-multiplicity pp collisions, the lifetime of the hadronic phase is expected to be very short~\cite{ALICE:2023edr}, which may lead to behavior different from that observed in heavy-ion collisions, where rescattering effects are dominant. In such small systems, multiple effects, including resonance lifetime, regeneration-driven yield enhancement, and rescattering-induced signal suppression, can interact in nontrivial ways. Therefore, a systematic study involving multiple resonance species is essential to disentangle these effects.

Furthermore, by extending the study beyond pp collisions to include p--O and O--O systems, we aim to investigate the system-size (or multiplicity) dependence of resonance production in a continuous manner. This extension is particularly timely, given that the LHC delivered its first oxygen-beam collisions in 2025, providing experimental data in a system-size regime intermediate between pp/p–Pb and Pb–Pb collisions~\cite{ALICE:2026zck, CMS:2025bta}. The O--O system serves as a crucial bridge for a controlled system-size scan, as it occupies a unique regime of geometry and multiplicity that allows one to disentangle initial-state nuclear geometry from final-state medium dynamics. Early measurements from ALICE and CMS already indicate that certain QGP-like signatures, including flow-like collective behavior and evidence for parton energy loss, emerge in O--O collisions, suggesting that even these relatively small oxygen systems may sustain a transient hadronic phase. Whether such a phase is sufficiently long-lived to leave observable modifications in resonance yields remains an open and fundamental question. This approach provides a unified picture of how particle production evolves from small to larger systems and offers deeper insight into the interplay between resonance lifetimes and hadronic interactions.

\section{Simulation framework}  
\label{sec:Simul}

\subsection{EPOS4 and UrQMD}
\label{sec:epos4}

The EPOS model~\cite{Werner:2013tya, Werner:2023zvo} is a general-purpose event generator that incorporates (3+1)-dimensional viscous hydrodynamic evolution. In the initial stage, multiple scatterings are described in terms of pomerons and strings. The reaction volume is then conceptually divided into two components: the core and the corona~\cite{Werner:2007bf}. The core component provides the initial condition for the QGP evolution, which is modeled using viscous hydrodynamics, whereas the corona component consists of hadrons produced directly from string fragmentation. The UrQMD model~\cite{Bleicher:1999xi, Bass:1998ca} is a hadronic transport model that describes final-state interactions such as rescattering and regeneration. In the EPOS framework, after the hydrodynamic core hadronizes, the hadrons produced from both the core and corona components are propagated through UrQMD. During this hadronic stage, chemical and kinetic freeze-out occur sequentially: chemical freeze-out takes place shortly after hadronization, while kinetic freeze-out occurs later as the system continues to expand and cool.

By incorporating UrQMD into the EPOS framework, final-state hadronic interactions, including rescattering and regeneration, can be accounted for. Previous studies have shown that EPOS3 combined with UrQMD can describe resonance production in Pb–Pb collisions~\cite{Knospe:2015nva, ALICE:2019xyr}. In this study, we perform calculations for the resonance species listed in Table~\ref{tab:table_res} in pp collisions at $\sqrt{s}=13.6$ TeV, p--O collisions at $\sqrt{s}=9.62$ TeV, and O--O and Pb--Pb collisions at $\sqrt{s_{\mathrm{NN}}}=5.36$ TeV. By comparing results obtained with and without the UrQMD hadronic cascade, we investigate how hadronic interactions, such as rescattering and regeneration, modify resonance production yields.

\subsection{Analysis method}

In this analysis, simulations are performed using the EPOS4 v4.0.3 event generator. To ensure consistency with the event and particle selection criteria used in ALICE analyses, only events containing at least one primary charged particle within $|\eta| < 1.0$, corresponding to the $\mathrm{INEL}>0$ event class in pp collisions, are selected. The event multiplicity classification is based on the number of primary charged particles within the FT0M detector acceptance, defined as $-3.3 < \eta < -2.1$ and $3.5 < \eta < 4.9$. The analysis focuses on resonance particles and their corresponding stable-particle reference species, all of which are selected within the rapidity interval $|y| < 0.5$. The resonances are reconstructed using the decay channels listed in Table~\ref{tab:table_res}.

The resonance yields and $\langle p_{\mathrm{T}} \rangle$ are evaluated for three configurations, depending on how the UrQMD hadronic cascade is treated. In the configuration without UrQMD, denoted as \textit{UrQMD OFF}, resonances and the corresponding stable particles are identified independently, and their yields and $\langle p_{\mathrm{T}} \rangle$ are calculated for each multiplicity class. When UrQMD is enabled, two configurations are considered. The first, denoted as \textit{UrQMD reg}, corresponds to the total number of resonances produced after the hadronic cascade, accounting for hadronic interactions that may increase or decrease resonance yields. The second, denoted as \textit{UrQMD reg+res}, excludes resonances whose decay daughters undergo rescattering. This selection accounts for the loss of reconstructible resonance signals due to rescattering and therefore provides the closest model representation of the experimentally observable resonance signal. We followed the same analysis method used in Ref.~\cite{Knospe:2015nva}. In the following results, \textit{UrQMD OFF}, \textit{UrQMD reg}, and \textit{UrQMD reg+res} are shown as black, blue, and red bands, respectively.

\section{Results and Discussions}
\label{sec:Results_Disscussion}

\subsection{$p_{T}$-integrated yields}

       \begin{figure*}[htb]
        \includegraphics[width=0.99\textwidth]{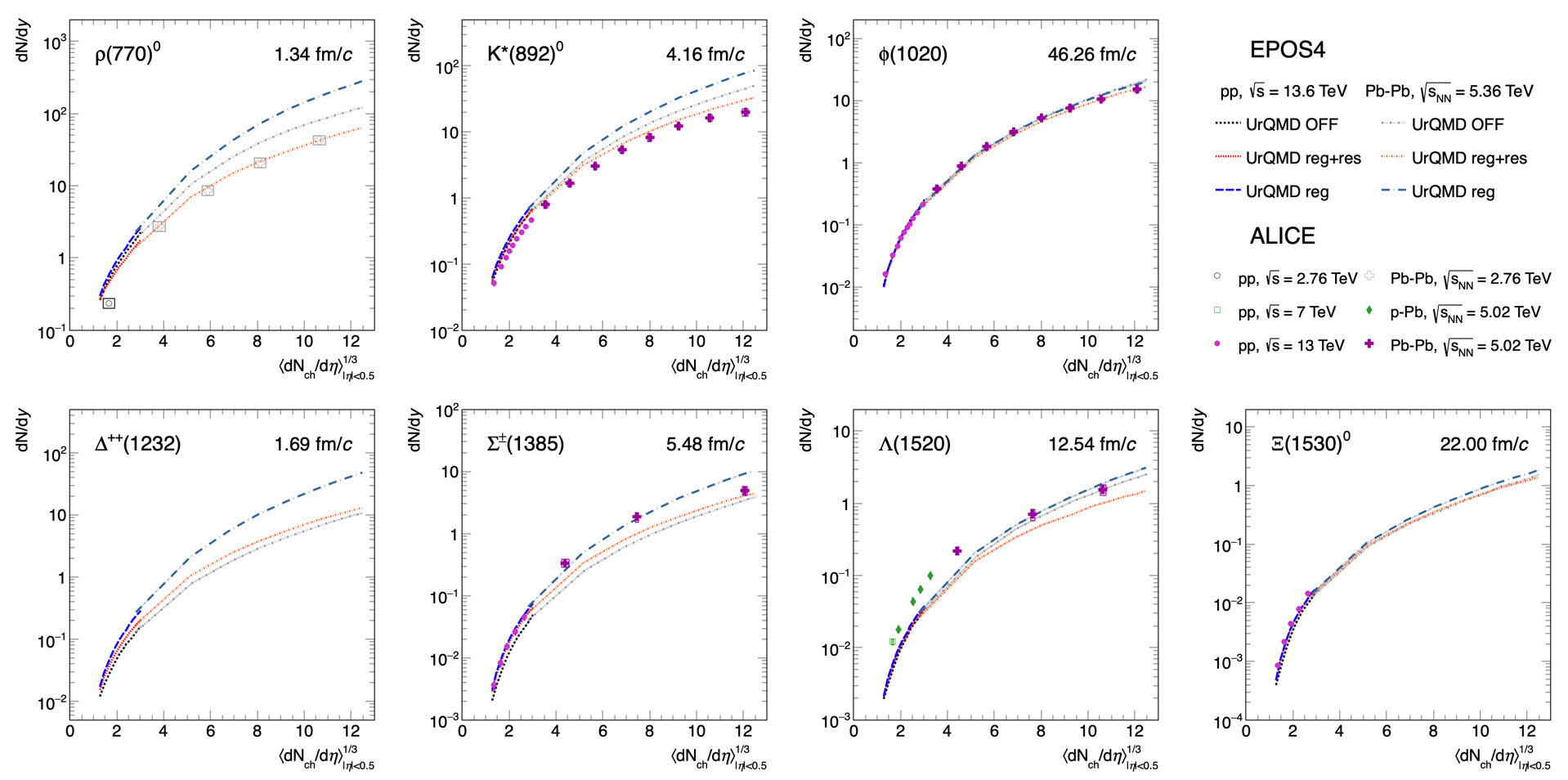}
        
        \caption{$p_{\mathrm{T}}$-integrated yields of various resonances as a function of $\langle dN_{\mathrm{ch}}/d\eta \rangle_{|\eta|<0.5}$ in pp collisions at $\sqrt{s}=13.6$ TeV and Pb--Pb collisions at $\sqrt{s_{\mathrm{NN}}}=5.36$ TeV, obtained from EPOS4 with three different hadronic-cascade configurations: UrQMD OFF, UrQMD reg+res, and UrQMD reg. The model results are compared with available experimental measurements at different collision energies.}
        \label{fig:fig_dNdy}
        \end{figure*}

        \begin{figure*}[htb]
        \includegraphics[width=0.99\textwidth]{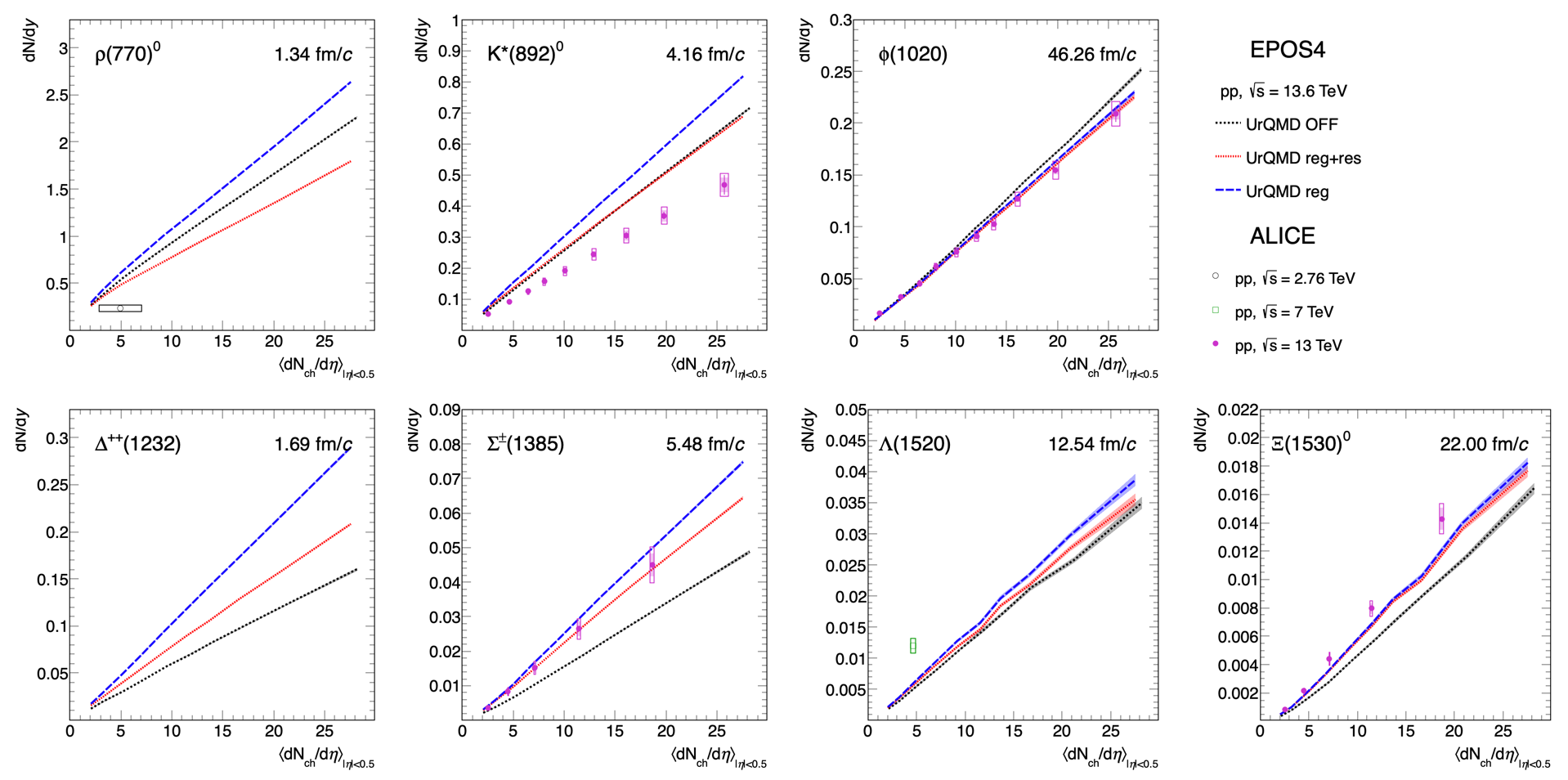}
        
        \caption{$p_{\mathrm{T}}$-integrated yields of various resonances as a function of $\langle dN_{\mathrm{ch}}/d\eta \rangle_{|\eta|<0.5}$ in pp collisions at $\sqrt{s}=13.6$ TeV, obtained from EPOS4 with three different hadronic-cascade configurations: UrQMD OFF, UrQMD reg+res, and UrQMD reg. The horizontal axis is shown on a linear scale.}
        \label{fig:fig_dNdy2}
        \end{figure*}

Figure~\ref{fig:fig_dNdy} shows the $p_{\mathrm{T}}$-integrated yields of various resonances as a function of the charged-particle multiplicity density at midrapidity, $\langle dN_{\mathrm{ch}}/d\eta \rangle_{|\eta|<0.5}$, obtained from the EPOS4 model. The results cover the multiplicity ranges corresponding to pp collisions at $\sqrt{s}=13.6$ TeV and Pb–Pb collisions at $\sqrt{s_{\mathrm{NN}}}=5.36$ TeV. Three hadronic-cascade configurations are compared: UrQMD OFF, UrQMD reg+res, and UrQMD reg. The model results are also compared with available experimental measurements at different collision energies.

In general, the UrQMD reg configuration, which includes resonance production via hadronic interactions in the cascade, yields more resonances than the UrQMD OFF configuration. This indicates that regeneration processes in the hadronic phase can increase the total number of resonances produced. When the rescattering of resonance decay daughters is taken into account, as in the UrQMD reg+res configuration, the reconstructible resonance yields are reduced. This reduction reflects the loss of experimentally observable resonance signals due to rescattering of daughter particles in the hadronic phase.

The relative difference between the three configurations depends strongly on the resonance species. For short-lived resonances such as $\rho(770)^0$ and $\mathrm{K}^{*}(892)^0$, the UrQMD reg+res yields are lower than the UrQMD reg yields, demonstrating the importance of rescattering effects. For longer-lived resonances such as $\phi(1020)$, the difference between the configurations is smaller, consistent with the expectation that such particles are less affected by the hadronic phase. For some baryonic resonances, the enhancement from regeneration can be comparable to or greater than the reduction due to rescattering, leading to a different ordering among the UrQMD configurations. These results show that the final resonance yield is governed by the competition between regeneration and rescattering, rather than by the resonance lifetime alone.

The comparison with experimental data indicates that EPOS4 with the UrQMD reg+res configuration provides a reasonable description of the measured yields for several mesonic resonances, including $\rho(770)^0$, $\mathrm{K}^{*}(892)^0$, and $\phi(1020)$, within the available multiplicity range. For some baryonic resonances, however, the model tends to underestimate the measured yields. This suggests that the description of baryonic resonance production, including the balance between primary production, regeneration, and rescattering, may require further tuning or additional constraints from data.

Figure~\ref{fig:fig_dNdy2} focuses on pp collisions at $\sqrt{s}=13.6$ TeV, with the horizontal axis shown on a linear scale. This representation allows the multiplicity dependence in the pp region to be examined more clearly. In this smaller system, the separation between the UrQMD OFF and UrQMD reg+res configurations is generally smaller than in the Pb–Pb region, reflecting the shorter duration and smaller volume of the hadronic phase. Nevertheless, visible differences among the configurations remain for several resonance species, indicating that hadronic interactions can still modify resonance production even in pp collisions.

The pp-focused comparison also shows that the relative importance of regeneration and rescattering differs by particle species. For $\mathrm{K}^{*}(892)^0$, the difference between UrQMD OFF and UrQMD reg+res becomes small over much of the pp multiplicity range, suggesting that regeneration and rescattering effects partially compensate each other. In contrast, for $\rho(770)^0$, which has a shorter lifetime than $\mathrm{K}^{*}(892)^0$, the UrQMD reg+res yield remains lower than the UrQMD OFF yield. This indicates that for $\rho(770)^0$, the loss of reconstructible resonance signals due to rescattering exceeds the yield enhancement from regeneration. Some baryonic resonances exhibit a different ordering between configurations, indicating a stronger role for regeneration-driven yield enhancement. These observations support the interpretation that, in small collision systems, resonance production is influenced not only by lifetime-dependent rescattering but also by regeneration and species-dependent hadronic interaction dynamics.


\subsection{Particle yield ratio}

        \begin{figure*}[htb]
        \includegraphics[width=0.99\textwidth]{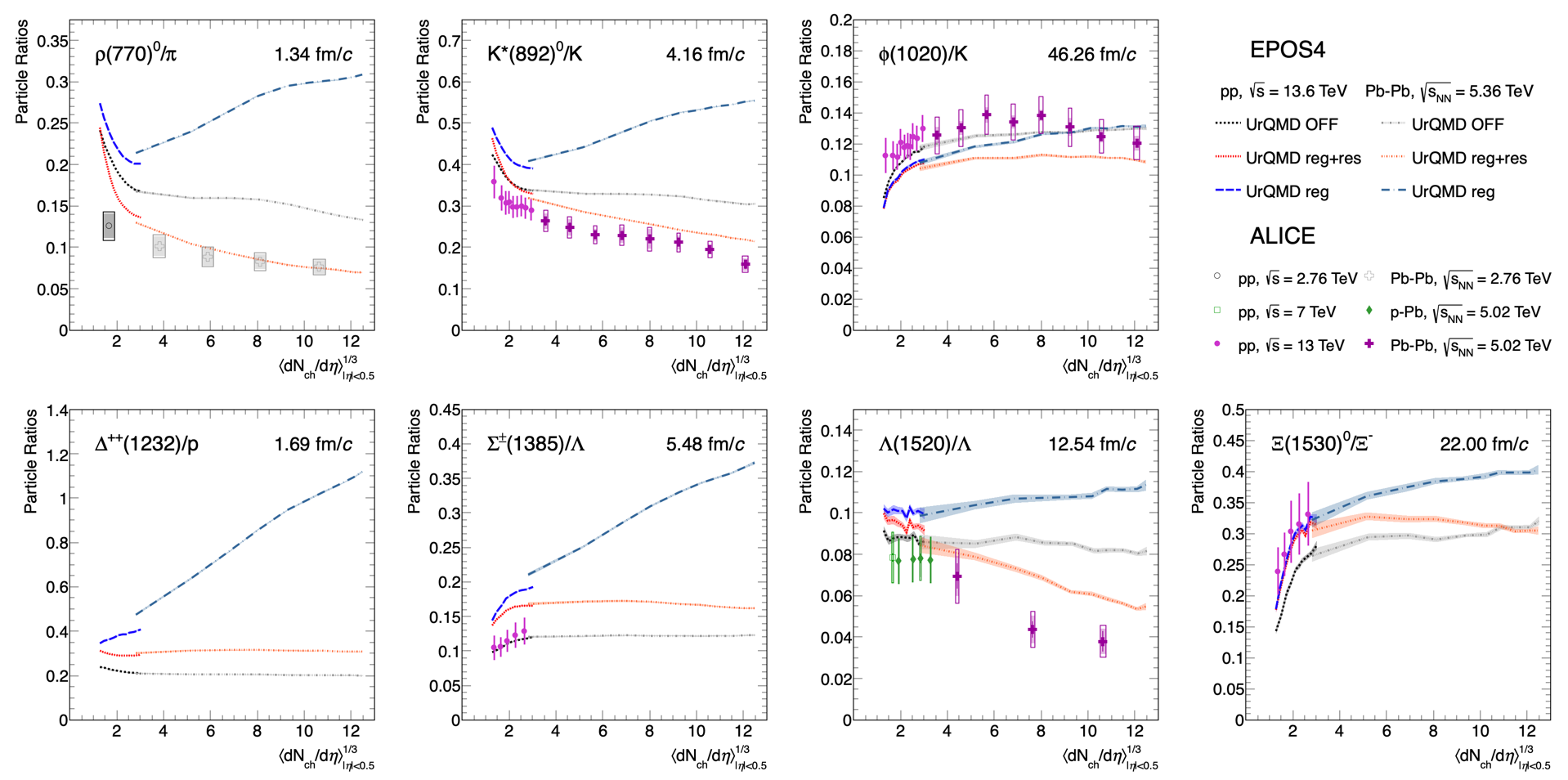}
        \caption{Particle yield ratios of various resonances to their corresponding ground-state particles as a function of $\langle dN_{\mathrm{ch}}/d\eta \rangle_{|\eta|<0.5}$ in pp collisions at $\sqrt{s}=13.6$ TeV and Pb--Pb collisions at $\sqrt{s_{\mathrm{NN}}}=5.36$ TeV, obtained from EPOS4 with three hadronic-cascade configurations: UrQMD OFF, UrQMD reg+res, and UrQMD reg. The model results are compared with available experimental measurements at different collision energies.}
        \label{fig:fig_multRatio}
        \end{figure*}

         \begin{figure*}[htb]
        \includegraphics[width=0.99\textwidth]{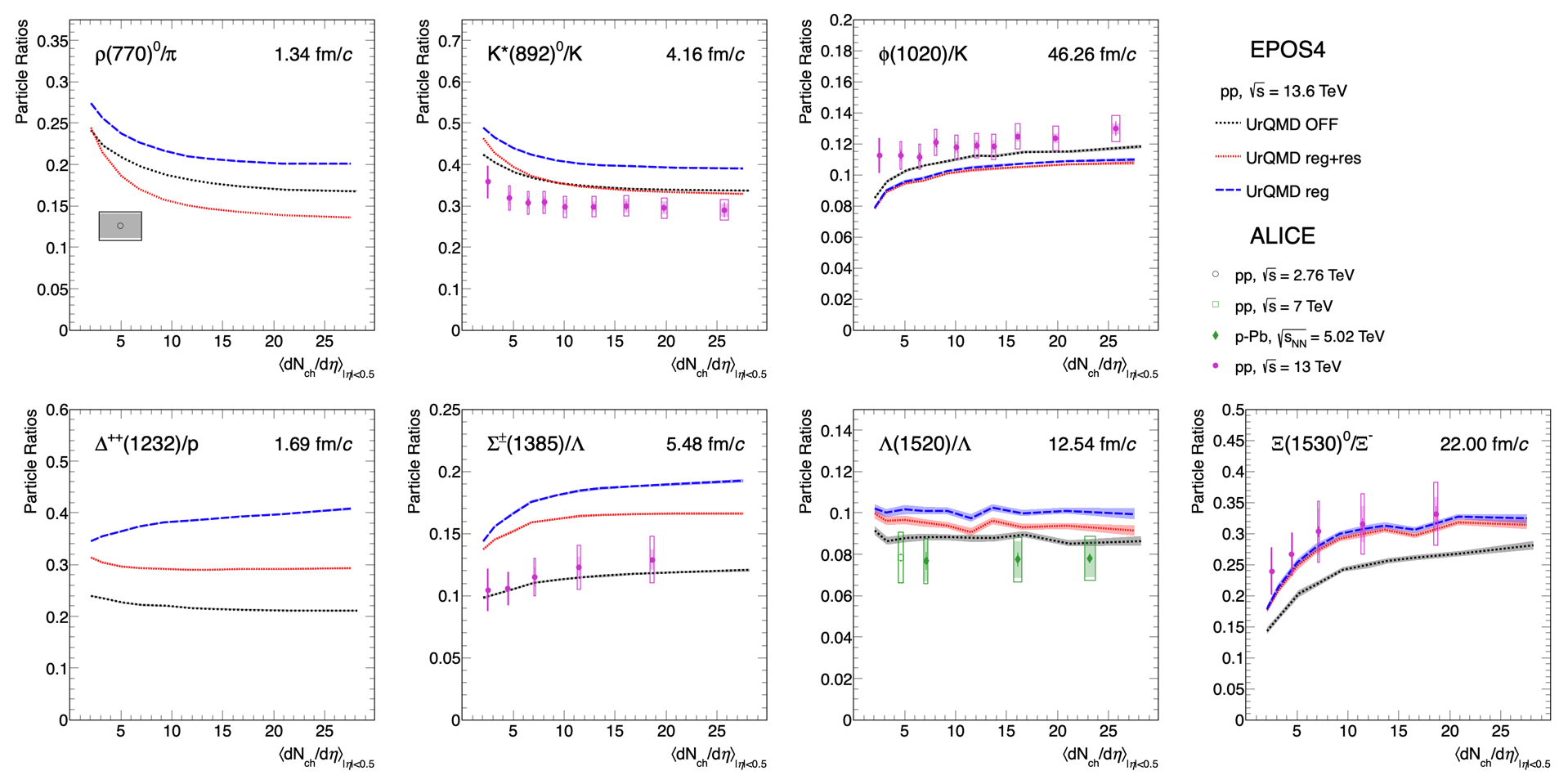}
        \caption{Particle yield ratios of various resonances to their corresponding ground-state particles as a function of $\langle dN_{\mathrm{ch}}/d\eta \rangle_{|\eta|<0.5}$ in pp collisions at $\sqrt{s}=13.6$ TeV, obtained from EPOS4 with three different hadronic-cascade configurations: UrQMD OFF, UrQMD reg+res, and UrQMD reg. The horizontal axis is shown on a linear scale.}
        \label{fig:fig_multRatio2}
        \end{figure*}

The yield ratios of resonances to their corresponding ground-state particles are key observables for probing hadronic-phase effects. If hadronic interactions in the hadronic phase do not significantly modify the production of either the resonance or its corresponding ground-state particle, the ratio is expected to remain approximately constant as a function of charged-particle multiplicity. Deviations from such behavior, therefore, provide sensitivity to rescattering and regeneration effects.

Figure~\ref{fig:fig_multRatio} shows the resonance-to-ground-state particle yield ratios as a function of $\langle dN_{\mathrm{ch}}/d\eta \rangle_{|\eta|<0.5}^{1/3}$, covering the multiplicity ranges corresponding to pp and Pb–Pb collisions. For short-lived mesonic resonances such as $\rho(770)^0$ and $\mathrm{K}^{*}(892)^0$, the UrQMD reg+res configuration reproduces the decreasing trend observed in the experimental data. This indicates that the loss of reconstructible resonance signals due to daughter-particle rescattering is essential for describing the measured multiplicity dependence. A similar behavior is observed for the $\Lambda(1520)/\Lambda$ ratios, though the effect is less pronounced for longer-lived resonances.

The comparison among the three UrQMD configurations shows that the magnitude of rescattering effects generally decreases toward smaller systems, consistent with the shorter lifetime and smaller volume of the hadronic phase. However, the observed behavior cannot be understood solely as a reduction of rescattering. In several cases, especially for baryonic resonances, the yield enhancement from regeneration is comparable to or greater than the suppression from rescattering. This demonstrates that the final resonance-to-ground-state particle ratios are governed by the competition between rescattering-induced signal loss and regeneration-driven yield enhancement.

For $\Delta^{++}(1232)/p$ and $\Sigma^{\pm}(1385)/\Lambda$, the UrQMD reg+res results are higher than those obtained with UrQMD OFF over the full multiplicity range. This indicates that, for these resonances, regeneration effects dominate over the loss of reconstructible signals due to rescattering. Such behavior highlights the importance of regeneration, which can play a significant role depending on the particle species, lifetime, and available hadronic interaction channels.

Figure~\ref{fig:fig_multRatio2} focuses on the pp collision region, with the horizontal axis shown on a linear scale. In this representation, the differences among the UrQMD configurations can be examined more clearly in small systems. For long-lived resonances such as $\phi(1020)$ and $\Xi(1530)^0$, the UrQMD reg and UrQMD reg+res results are very similar, indicating that daughter-particle rescattering has only a small impact. For other resonances, the separation between the configurations is also reduced compared with that in the Pb–Pb multiplicity region. Therefore, in pp collisions, rescattering effects are expected to be visible mainly for very short-lived resonances such as $\rho(770)^0$, consistent with the expectation that the hadronic phase lifetime in high-multiplicity pp collisions is short, on the order of $\sim 2~\mathrm{fm}/c$~\cite{ALICE:2023edr}.

At the same time, the pp results indicate that hadronic-phase effects in small systems should not be interpreted simply as a suppression mechanism. Even when rescattering effects are weak, regeneration can still modify resonance production, leading to species-dependent changes in yield ratios. In particular, the systematic reduction in the $\phi(1020)$ yield when UrQMD is enabled suggests that additional mechanisms, such as nuclear absorption~\cite{Gubler:2024day, Steinheimer:2025mho, HADES:2018qkj}, may influence the final resonance yield, although a detailed interpretation of this behavior requires further study. 
Since experimental measurements in small collision systems remain limited and the differences among model configurations can be subtle, more precise data will be essential to constrain the relative roles of rescattering and regeneration.

\subsection{Mean $p_{T}$}

        \begin{figure*}[htb]
        \includegraphics[width=0.99\textwidth]{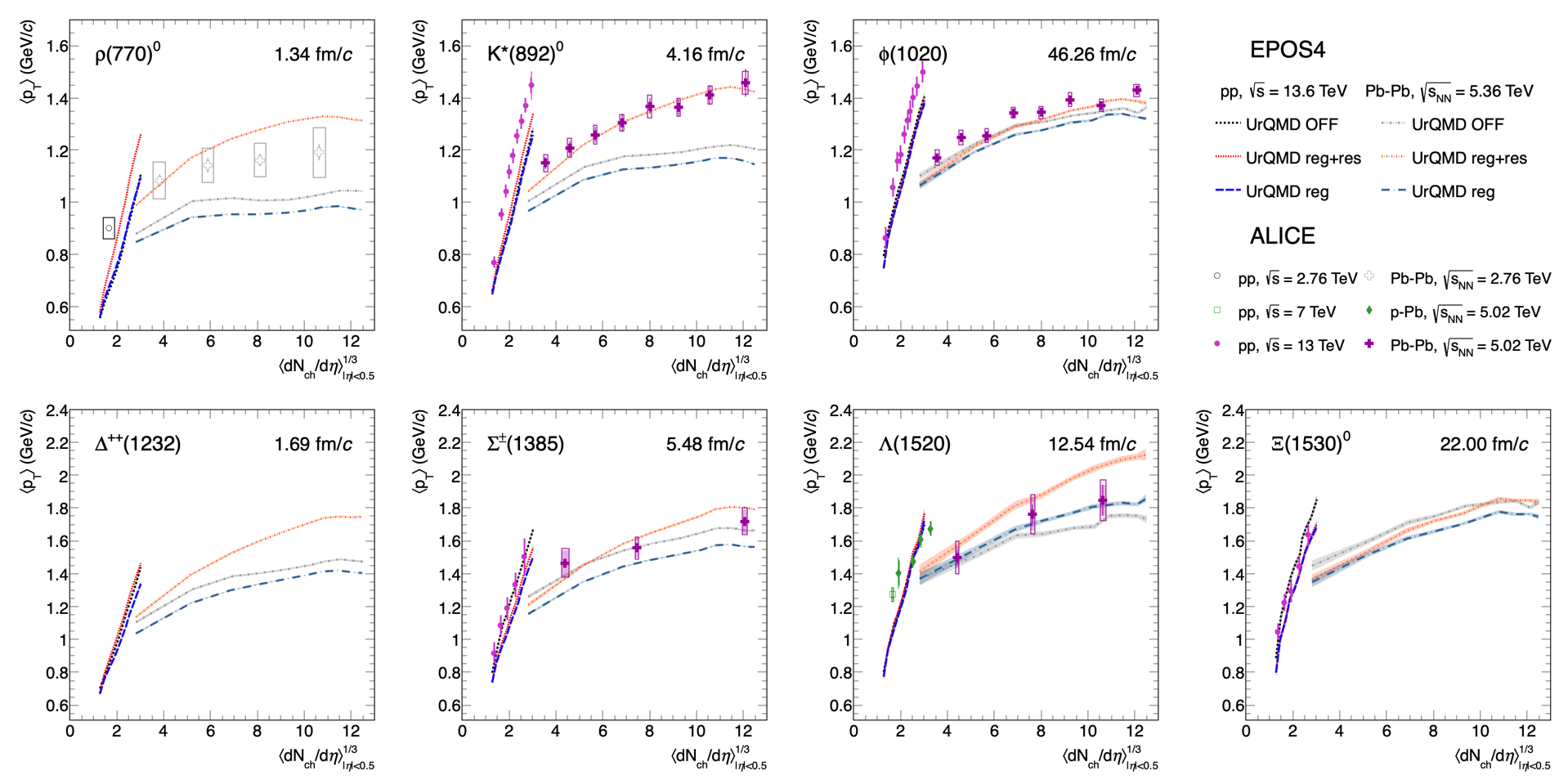}
        \caption{$\langle p_{\mathrm{T}} \rangle$ of various resonances as a function of $\langle dN_{\mathrm{ch}}/d\eta \rangle_{|\eta|<0.5}$ in pp collisions at $\sqrt{s}=13.6$ TeV and Pb--Pb collisions at $\sqrt{s_{\mathrm{NN}}}=5.36$ TeV, obtained from EPOS4 with three different hadronic-cascade configurations: UrQMD OFF, UrQMD reg+res, and UrQMD reg. The model results are compared with available experimental measurements at different collision energies.}
        \label{fig:fig_pT}
        \end{figure*}

        \begin{figure*}[htb]
        \includegraphics[width=0.99\textwidth]{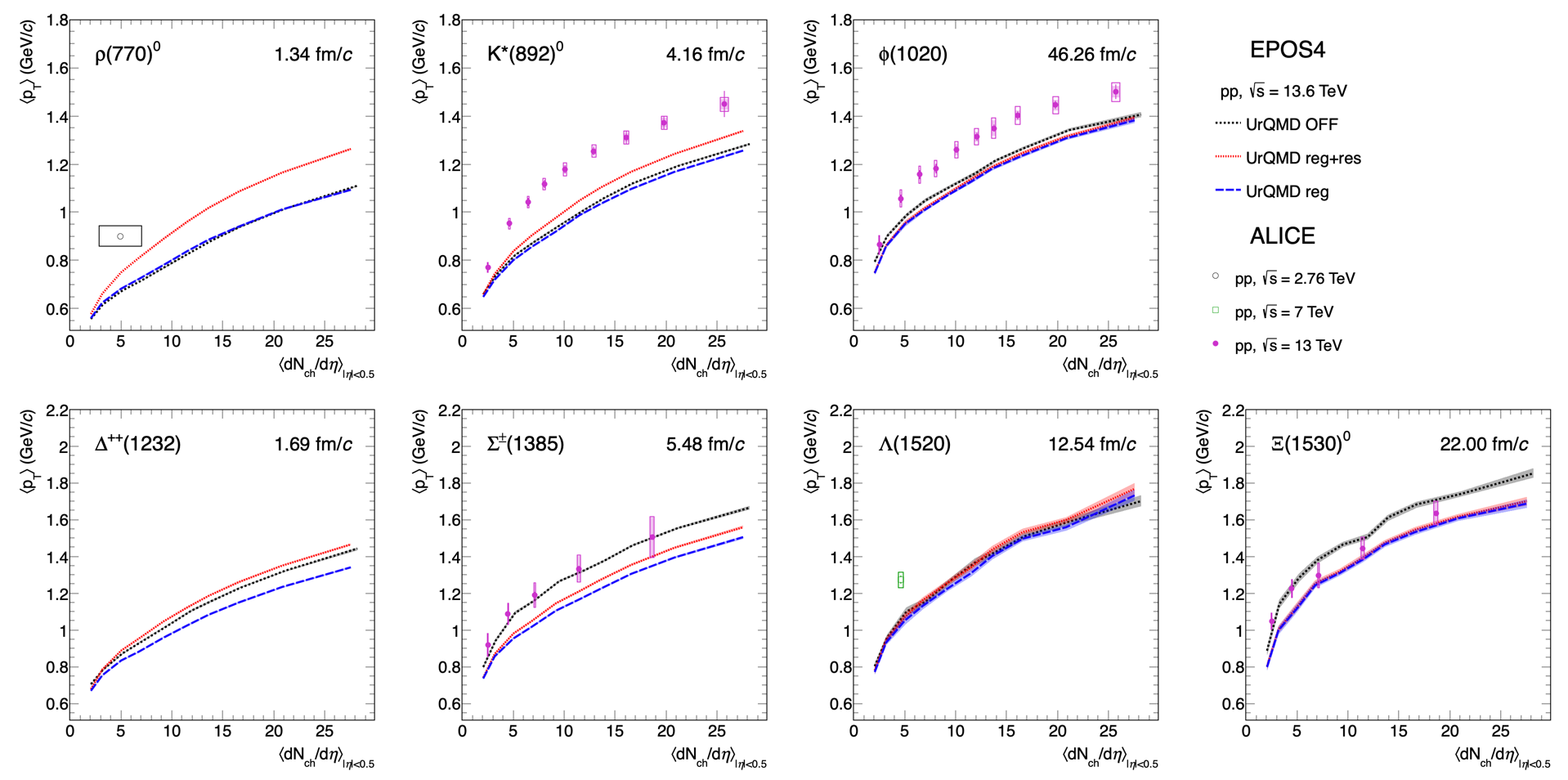}
        \caption{$\langle p_{\mathrm{T}} \rangle$ of various resonances as a function of $\langle dN_{\mathrm{ch}}/d\eta \rangle_{|\eta|<0.5}$ in pp collisions at $\sqrt{s}=13.6$ TeV, obtained from EPOS4 with three different hadronic-cascade configurations: UrQMD OFF, UrQMD reg+res, and UrQMD reg. The horizontal axis is shown on a linear scale.}
        \label{fig:fig_pT2}
        \end{figure*}

The $\langle p_{\mathrm{T}} \rangle$ of resonances provides complementary information on hadronic-phase interactions beyond that obtained from $p_{\mathrm{T}}$-integrated yields. Low-$p_{\mathrm{T}}$ particles are more likely to interact in the hadronic medium, and rescattering of resonance decay daughters can preferentially suppress reconstructible resonance signals in the low-$p_{\mathrm{T}}$ region. As a result, rescattering tends to increase the measured $\langle p_{\mathrm{T}} \rangle$. In contrast, resonances regenerated during the hadronic phase are generally expected to populate the lower-$p_{\mathrm{T}}$ region, thereby reducing the average transverse momentum. The measured $\langle p_{\mathrm{T}} \rangle$ therefore reflects the competition between rescattering-induced signal loss and regeneration-driven yield enhancement.

Figure~\ref{fig:fig_pT} shows the multiplicity dependence of $\langle p_{\mathrm{T}} \rangle$ for various resonances in pp and Pb--Pb collisions. In general, the UrQMD reg configuration gives lower $\langle p_{\mathrm{T}} \rangle$ values than the UrQMD OFF configuration, indicating that resonance regeneration in the hadronic cascade enhances the low-$p_{\mathrm{T}}$ component of the final-state resonance population. When the rescattering of decay daughters is included, as in the UrQMD reg+res configuration, $\langle p_{\mathrm{T}} \rangle$ increases relative to the UrQMD reg case. This behavior is consistent with the preferential loss of reconstructible resonance signals at low $p_{\mathrm{T}}$ due to rescattering of daughter particles.

As observed for the integrated yields, the relative importance of regeneration and rescattering depends strongly on the resonance species and the collision system. For some resonances, regeneration dominates over rescattering, leading to a sizable reduction of $\langle p_{\mathrm{T}} \rangle$ compared with the UrQMD OFF case. For other resonances, the increase due to rescattering partially offsets the regeneration effect. Consequently, the relative ordering of the UrQMD OFF, UrQMD reg, and UrQMD reg+res configurations varies with particle species and charged-particle multiplicity.

Figure~\ref{fig:fig_pT2} focuses on the pp collision region, with the horizontal axis on a linear scale. Compared with the Pb--Pb multiplicity region, the overall separation among the three configurations becomes smaller, reflecting the reduced volume and shorter lifetime of the hadronic phase in pp collisions. Nevertheless, visible differences remain for several resonance species, indicating that hadronic interactions can still modify the transverse-momentum distributions even in small collision systems.

\begin{figure*}[htb]
        \includegraphics[width=0.99\textwidth]{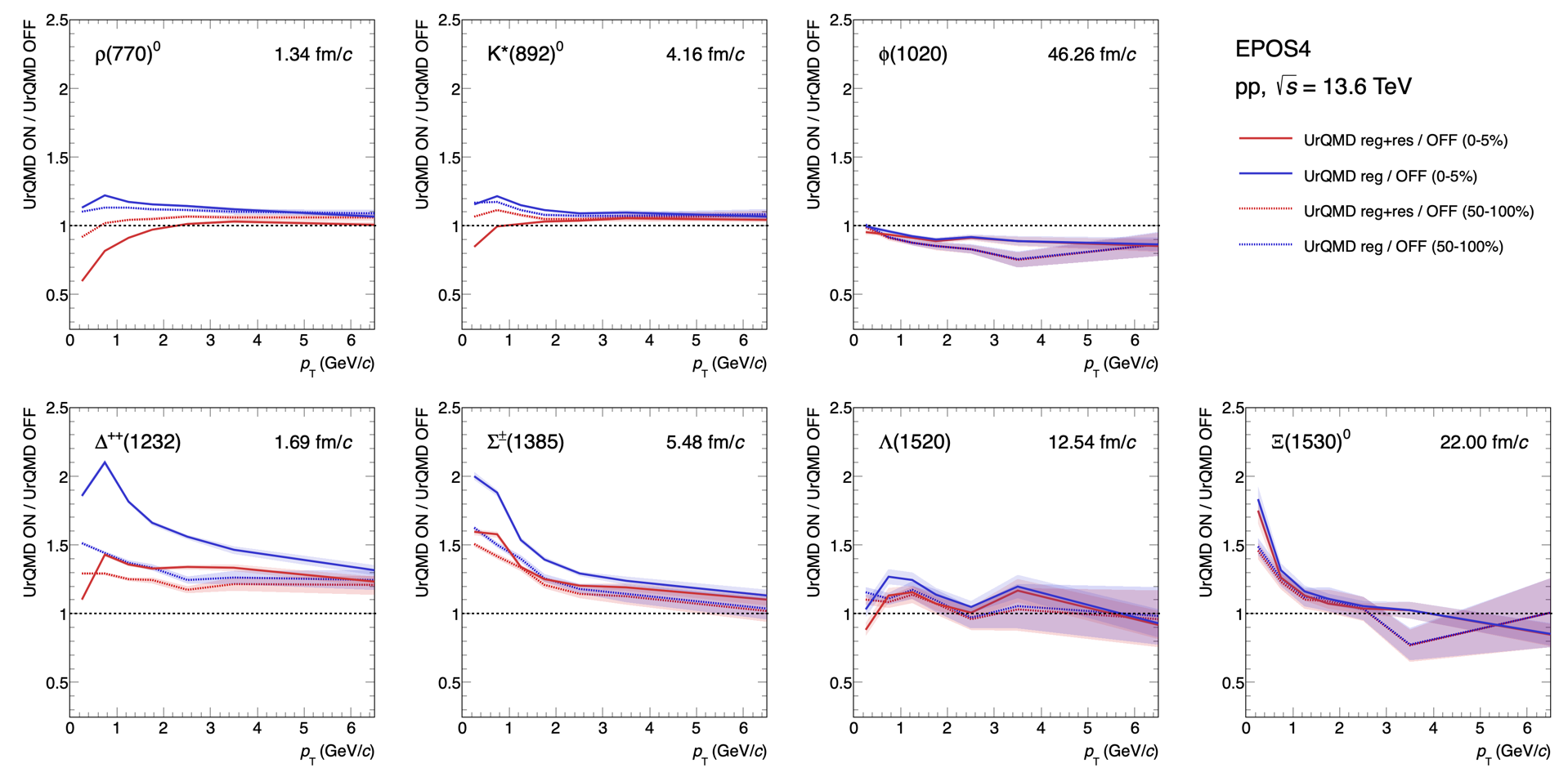}
        \caption{Ratio of UrQMD ON to UrQMD OFF yields as a function of $p_{\mathrm{T}}$ for various resonances in pp collisions at $\sqrt{s}=13.6$ TeV. The red curves represent the UrQMD reg+res/UrQMD OFF ratio, while the blue curves represent the UrQMD reg/UrQMD OFF ratio. Solid and dashed curves correspond to the 0--5\% high-multiplicity and 50--100\% low-multiplicity classes, respectively.}
        \label{fig:fig_pTratio}
\end{figure*}

Figure~\ref{fig:fig_pTratio} provides further insight into the mechanisms underlying the observed modifications in $\langle p_{\mathrm{T}} \rangle$. It shows the ratio of the UrQMD ON to UrQMD OFF results as a function of $p_{\mathrm{T}}$ in pp collisions at $\sqrt{s}=13.6$ TeV. The ratios are shown separately for the UrQMD reg and UrQMD reg+res configurations in the 0--5\% high-multiplicity and 50--100\% low-multiplicity classes.

As expected, the UrQMD reg configuration shows an enhancement of resonance production relative to UrQMD OFF over a broad $p_{\mathrm{T}}$ range due to regeneration in the hadronic cascade. This enhancement is generally most pronounced at low $p_{\mathrm{T}}$, indicating that regenerated resonances are preferentially produced with relatively small transverse momentum. As a result, the regeneration effect tends to reduce $\langle p_{\mathrm{T}} \rangle$ values. Since the enhancement extends over a finite $p_{\mathrm{T}}$ range, however, its impact on $\langle p_{\mathrm{T}} \rangle$ can be more moderate than its effect on the $p_{\mathrm{T}}$-integrated yields.

The effect of daughter-particle rescattering exhibits a different behavior. When rescattering is taken into account, the loss of reconstructible resonance signals is concentrated mainly in the low-$p_{\mathrm{T}}$ region. This preferential suppression of low-$p_{\mathrm{T}}$ resonances increases the measured $\langle p_{\mathrm{T}} \rangle$. The magnitude of this effect depends on the multiplicity class: the difference between the UrQMD reg and UrQMD reg+res configurations is considerably larger in the 0--5\% high-multiplicity class than in the 50--100\% low-multiplicity class. This indicates that rescattering becomes more important as the size and lifetime of the hadronic phase increase.

The transverse-momentum dependence of regeneration also shows a strong particle-species dependence. Resonances such as $\Delta^{++}(1232)$ and $\Sigma^{\pm}(1385)$, for which regeneration gives a sizable contribution to the yield modification, exhibit a pronounced enhancement in the low-$p_{\mathrm{T}}$ region. In contrast, particles such as $\rho(770)^{0}$, $\mathrm{K}^{*}(892)^{0}$, and $\Lambda(1520)$ show a broader and more uniform regeneration contribution over $p_{\mathrm{T}}$. Consequently, regeneration can modify $\langle p_{\mathrm{T}} \rangle$ differently depending on the particle species and may even lead to an increase in $\langle p_{\mathrm{T}} \rangle$ in some cases. This behavior indicates that regeneration is governed not only by the resonance lifetime but also by species-dependent hadronic interaction channels.

Overall, these results demonstrate that the modification of transverse-momentum distributions is governed by the interplay between regeneration and rescattering. Rescattering shows a relatively clear dependence on the resonance lifetime and becomes more significant in higher-multiplicity events, where the hadronic phase is expected to be larger and longer-lived. Regeneration, on the other hand, exhibits a stronger particle-species dependence, both in magnitude and in its $p_{\mathrm{T}}$ distribution. Therefore, the observed $\langle p_{\mathrm{T}} \rangle$ cannot be understood solely in terms of resonance lifetime or system size. Instead, species-dependent hadronic interaction dynamics and the balance between regeneration and rescattering must be considered simultaneously. Together with the integrated yields and resonance-to-ground-state particle ratios, the study of $\langle p_{\mathrm{T}} \rangle$ provides complementary information for a more complete understanding of hadronic-phase dynamics.

\subsection{p--O and O--O collision}

        \begin{figure*}[htb]
        \includegraphics[width=0.99\textwidth]{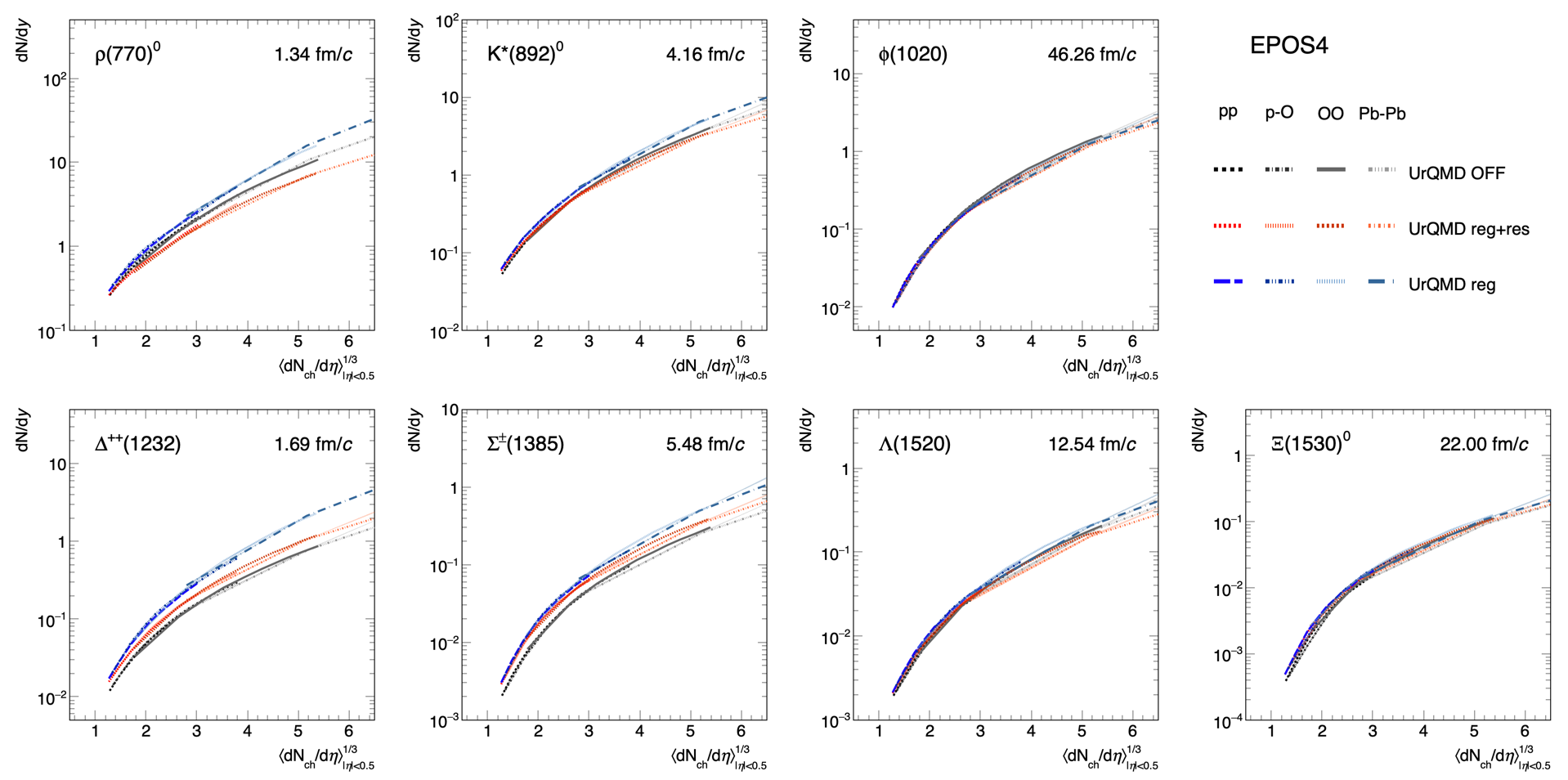}
        
        \caption{$p_{\mathrm{T}}$-integrated yields as a function of $\langle dN_{\mathrm{ch}}/d\eta \rangle_{|\eta|<0.5}^{1/3}$ in pp, p--O, O--O, and Pb--Pb collisions, obtained from EPOS4 with three hadronic-cascade configurations: UrQMD OFF, UrQMD reg+res, and UrQMD reg. The results are compared across collision systems to investigate the system-size dependence of resonance production.}
        \label{fig:fig_dNdy3}
        \end{figure*}

        \begin{figure*}[htb]
        \includegraphics[width=0.99\textwidth]{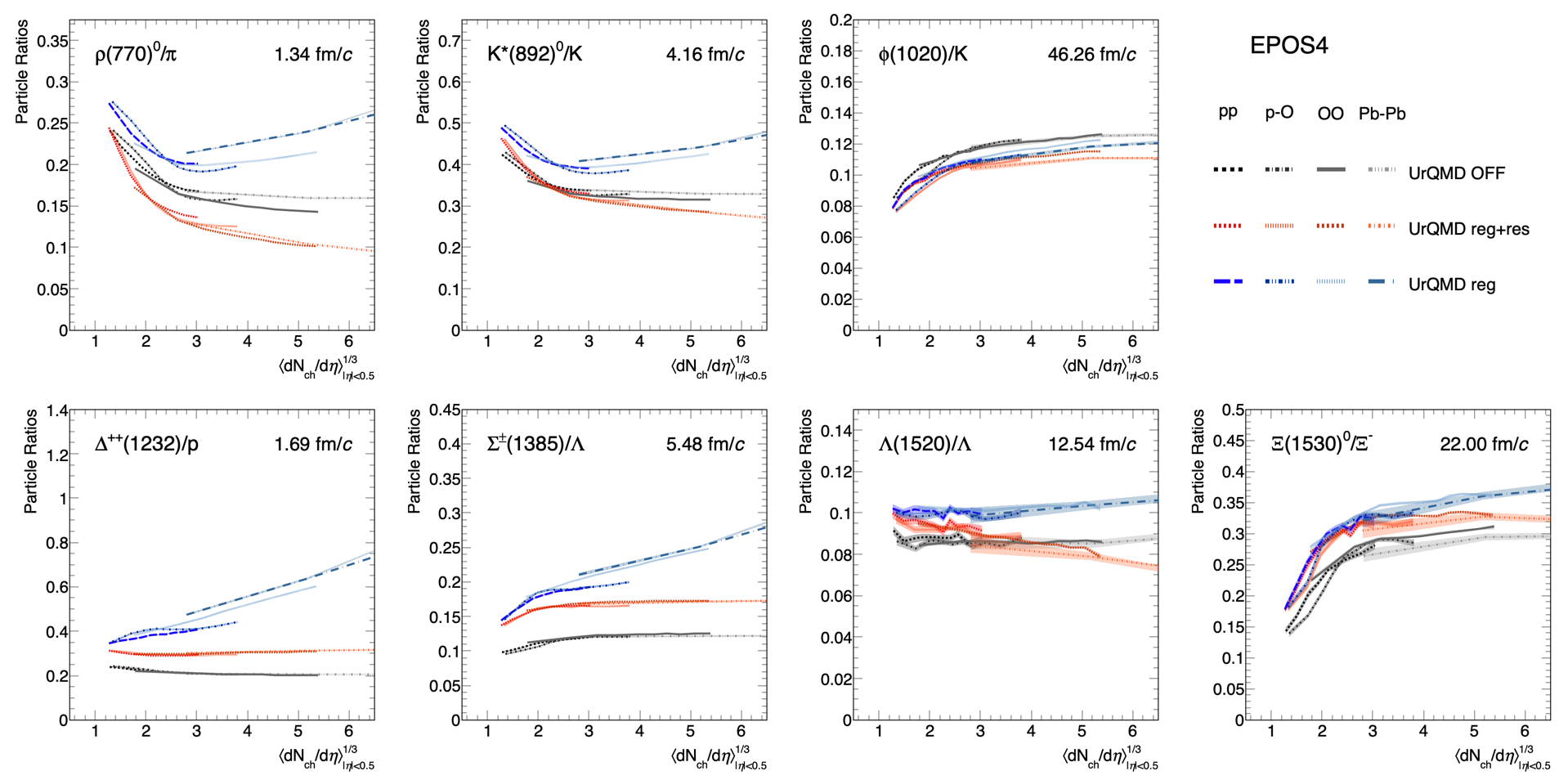}
        \caption{Particle yield ratios of various resonances to their corresponding ground-state particles as a function of $\langle dN_{\mathrm{ch}}/d\eta \rangle_{|\eta|<0.5}^{1/3}$ in pp, p--O, O--O, and Pb--Pb collisions, obtained from EPOS4 with three hadronic-cascade configurations: UrQMD OFF, UrQMD reg+res, and UrQMD reg. The results are compared across collision systems to investigate the system-size dependence of resonance production.}
        \label{fig:fig_multRatio3}
        \end{figure*}

        \begin{figure*}[htb]
        \includegraphics[width=0.99\textwidth]{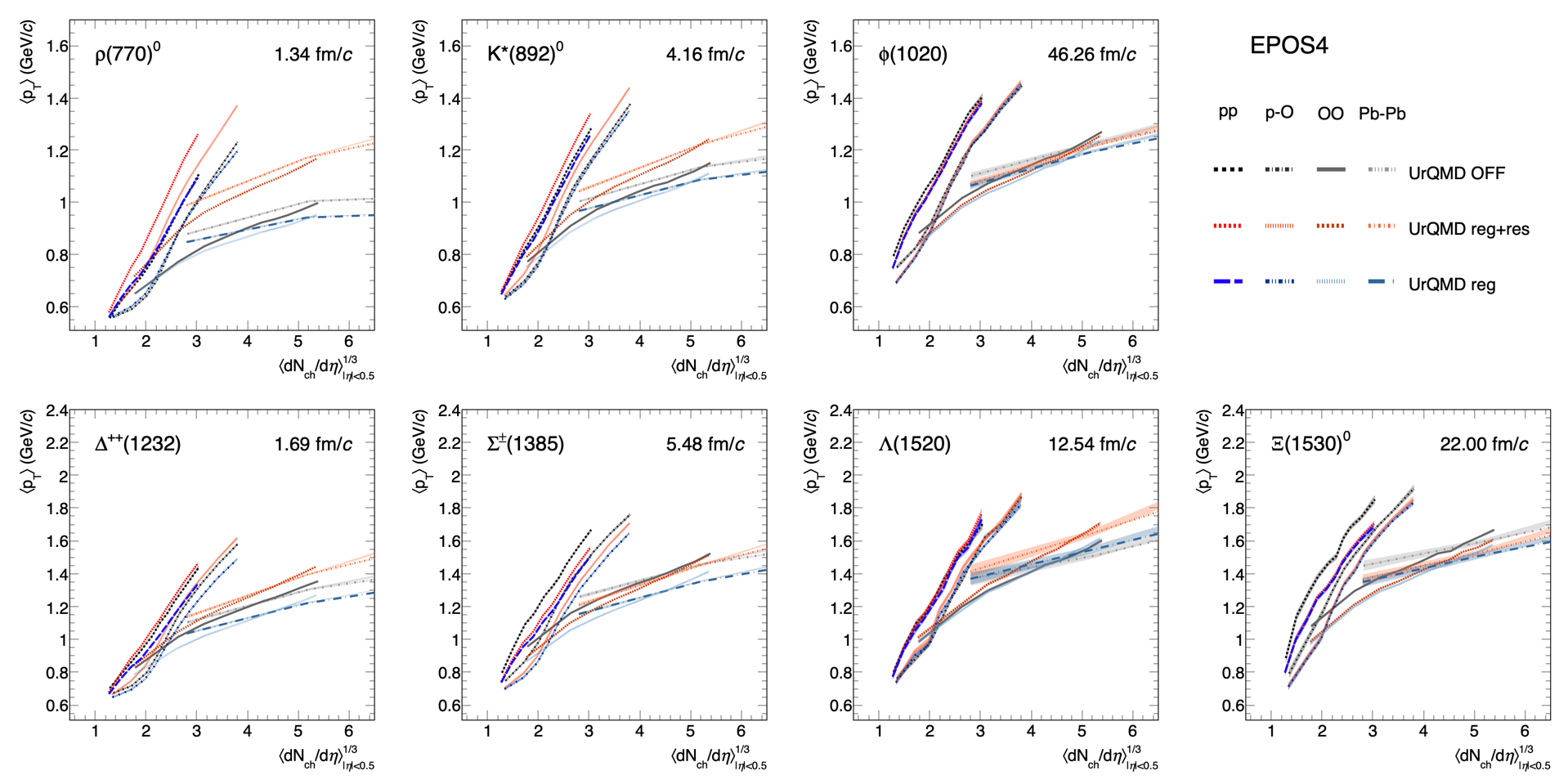}
        \caption{$\langle p_{\mathrm{T}} \rangle$ of various resonances as a function of $\langle dN_{\mathrm{ch}}/d\eta \rangle_{|\eta|<0.5}^{1/3}$ in pp, p--O, O--O, and Pb--Pb collisions, obtained from EPOS4 with three hadronic-cascade configurations: UrQMD OFF, UrQMD reg+res, and UrQMD reg. The results are compared across collision systems to investigate the system-size dependence of resonance production.}
        \label{fig:fig_pT3}
        \end{figure*}

Figures~\ref{fig:fig_dNdy3}, \ref{fig:fig_multRatio3}, and \ref{fig:fig_pT3} show the $p_{\mathrm{T}}$-integrated yields, resonance-to-ground-state particle ratios, and $\langle p_{\mathrm{T}} \rangle$, respectively, as a function of $\langle dN_{\mathrm{ch}}/d\eta \rangle_{|\eta|<0.5}^{1/3}$ for pp, p--O, O--O, and Pb--Pb collisions. The use of $\langle dN_{\mathrm{ch}}/d\eta \rangle^{1/3}$ provides a convenient proxy for the linear scale of the produced system and allows the system-size dependence of resonance production to be examined across a broad range of collision systems. Since p--O and O--O collisions populate the multiplicity region between pp and Pb–Pb collisions, they provide an important bridge for studying the continuous evolution of hadronic-phase effects from small to large systems.

As shown in Fig.~\ref{fig:fig_dNdy3}, the $p_{\mathrm{T}}$-integrated yields increase smoothly with $\langle dN_{\mathrm{ch}}/d\eta \rangle^{1/3}$ for all resonance species. The p--O and O--O results extend the trends observed in pp collisions toward higher multiplicities and connect them to the Pb–Pb region. The comparison among different collision systems indicates that, for the integrated yields, the overall multiplicity dependence is largely driven by the final-state charged-particle multiplicity rather than by the collision system alone. This suggests that total resonance production evolves approximately continuously with the size of the produced medium.

A similar behavior is observed for the resonance-to-ground-state particle ratios shown in Fig.~\ref{fig:fig_multRatio3}. The p--O and O--O results generally lie between the pp and Pb--Pb trends and follow a smooth multiplicity dependence. This indicates that the relative production of resonances with respect to their corresponding ground-state particles is also largely controlled by the produced multiplicity. At the same time, the separation among the UrQMD OFF, UrQMD reg+res, and UrQMD reg configurations demonstrates that hadronic interactions remain important throughout the full system-size range. The competition between regeneration and daughter-particle rescattering therefore evolves continuously from pp to Pb--Pb collisions, with p--O and O--O providing intermediate reference systems.

The behavior of $\langle p_{\mathrm{T}} \rangle$, shown in Fig.~\ref{fig:fig_pT3}, exhibits a more pronounced dependence on the collision system. While the p--O results tend to extend the pp trend to higher multiplicities, the O--O results are closer to the Pb--Pb behavior in the lower-multiplicity region. As a result, the $\langle p_{\mathrm{T}} \rangle$ values do not collapse onto a single universal curve as clearly as the integrated yields or yield ratios. This indicates that $\langle p_{\mathrm{T}} \rangle$ is sensitive not only to the final-state multiplicity but also to the underlying collision geometry and initial-state dynamics.

Such system dependence can be understood through the various mechanisms that generate similar final-state multiplicities across different collision systems. In high-multiplicity pp and p--O collisions, hard-scattering processes and jet-related contributions can play a relatively larger role, thereby enhancing the high-$p_{\mathrm{T}}$ component of the particle spectra and increasing $\langle p_{\mathrm{T}} \rangle$. In contrast, comparable multiplicities in O--O and Pb--Pb collisions are more likely to arise from a larger interaction volume and stronger collective expansion. Therefore, unlike the integrated yields and yield ratios, $\langle p_{\mathrm{T}} \rangle$ retains stronger sensitivity to the collision system and does not exhibit simple multiplicity scaling.

Nevertheless, the qualitative ordering among the UrQMD OFF, UrQMD reg, and UrQMD reg+res configurations remains broadly consistent across pp, p--O, O--O, and Pb--Pb collisions. The UrQMD reg configuration reflects the effect of resonance regeneration in the hadronic cascade, while the difference between UrQMD reg and UrQMD reg+res represents the loss of reconstructible resonance signals due to rescattering of daughter particles. The persistence of these differences in p--O and O--O collisions indicates that both regeneration and rescattering can affect resonance production even in intermediate-size systems. These results demonstrate that p--O and O--O collisions provide valuable experimental and theoretical reference systems for disentangling multiplicity-driven effects from genuine system-size and geometry-dependent dynamics.

\subsection{Estimation of the hadronic-phase lifetime}

\begin{figure*}[htb]
        \includegraphics[width=0.99\textwidth]{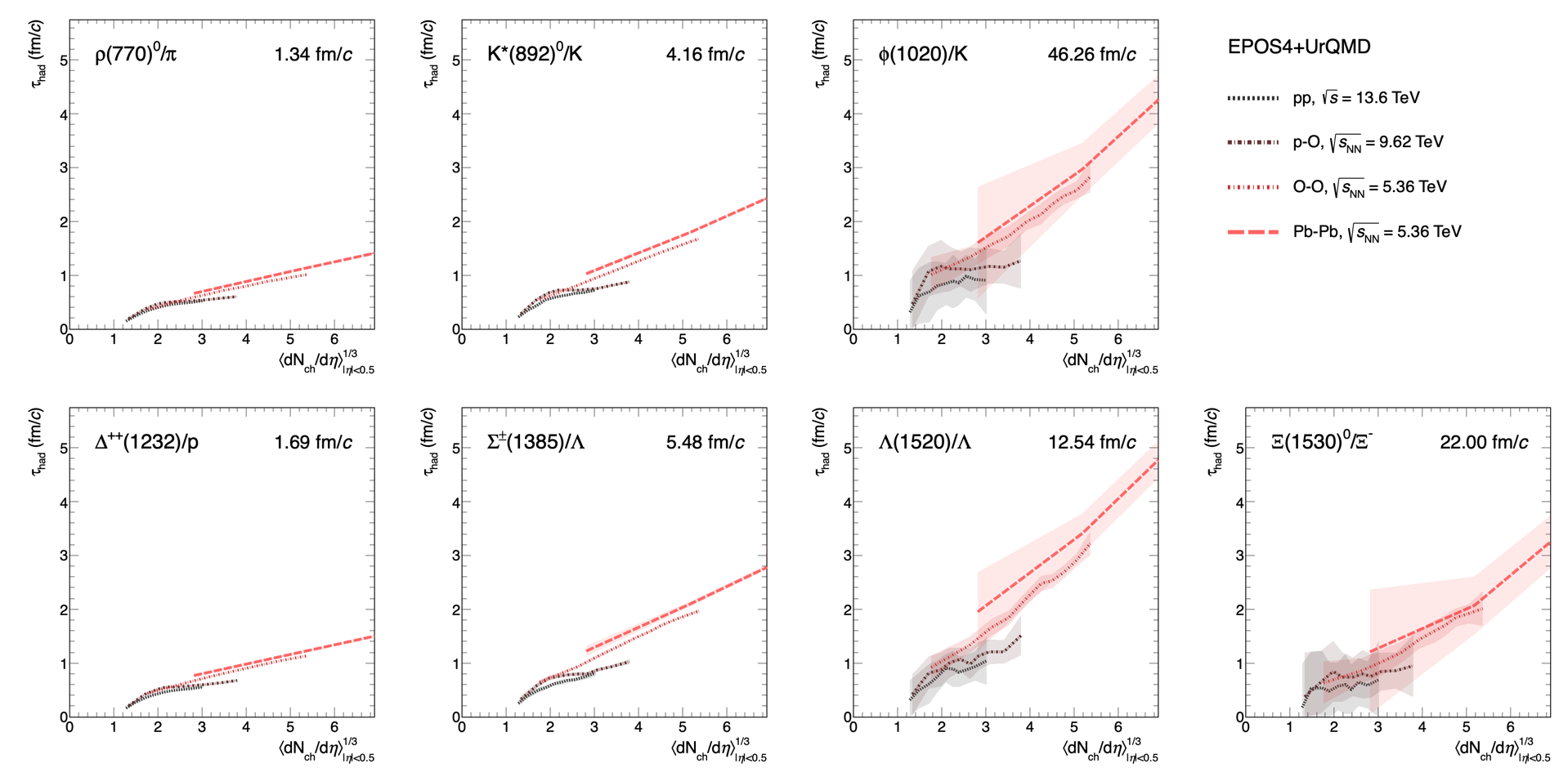}
        \caption{Estimated hadronic-phase lifetime, $\tau_{\mathrm{had}}$, as a function of $\langle dN_{\mathrm{ch}}/d\eta \rangle_{|\eta|<0.5}^{1/3}$ for pp, p--O, O--O, and Pb--Pb collisions. The hadronic-phase lifetime is extracted from the resonance-to-ground-state particle ratios obtained from the UrQMD reg and UrQMD reg+res configurations. The results are compared across collision systems to investigate the system-size dependence of the hadronic-phase lifetime.}
        \label{fig:fig_tau}
        \end{figure*}

In the previous sections, the effects of hadronic interactions on resonance production were investigated using $p_{\mathrm{T}}$-integrated yields, resonance-to-ground-state particle ratios, and $\langle p_{\mathrm{T}} \rangle$. These observables show that the relative importance of regeneration and rescattering depends strongly on both the resonance species and the collision system. In particular, the resonance-to-ground-state particle ratio is closely related to the duration of the hadronic phase, as it reflects both the loss of reconstructible resonance signals due to daughter-particle rescattering and the yield enhancement caused by regeneration. In this section, the hadronic-phase lifetime, $\tau_{\mathrm{had}}$, is estimated from the calculated ratios of resonance-to-ground-state particles.

Figure~\ref{fig:fig_tau} shows the estimated $\tau_{\mathrm{had}}$ as a function of $\langle dN_{\mathrm{ch}}/d\eta\rangle_{|\eta|<0.5}^{1/3}$ for pp, p--O, O--O, and Pb--Pb collisions. The quantity $\langle dN_{\mathrm{ch}}/d\eta\rangle^{1/3}$ serves as a proxy for the size of the produced system, enabling comparison of the system-size dependence of the hadronic-phase lifetime across different collision systems.

The hadronic-phase lifetime is commonly estimated from the change in the resonance-to-ground-state particle ratio between chemical and kinetic freeze-out. If a resonance decays during the hadronic phase and its decay daughters undergo rescattering, the original resonance signal may no longer be reconstructible. In this case, the ratio measured at kinetic freeze-out becomes lower than that at chemical freeze-out. In a simplified approximation in which regeneration is neglected, this reduction can be written as,
\begin{align}
\left[\frac{h^{*}}{h}\right]_{\mathrm{kinetic}}
=
\left[\frac{h^{*}}{h}\right]_{\mathrm{chemical}}
\exp\left(-\frac{\tau_{\mathrm{had}}}{\tau_{h^{*}}}\right),    
\end{align}
where $\left[h^{*}/h\right]_{\mathrm{chemical}}$ denotes the resonance-to-ground-state particle ratio at chemical freeze-out, and $\left[h^{*}/h\right]_{\mathrm{kinetic}}$ is the corresponding ratio at kinetic freeze-out. The quantity $\tau_{h^{*}}$ is the lifetime of the resonance in its rest frame, while $\tau_{\mathrm{had}}$ represents the time interval between chemical and kinetic freeze-out, namely the duration of the hadronic phase~\cite{ALICE:2019xyr, ALICE:2023edr}.

In experimental analyses, $\left[h^{*}/h\right]_{\mathrm{chemical}}$ cannot be measured directly and is therefore often approximated by the ratio measured in minimum-bias or low-multiplicity pp collisions, where hadronic-medium effects are assumed to be small. In the present EPOS4 study, however, the UrQMD hadronic cascade can be controlled within the model, allowing the resonance yields with and without rescattering effects to be compared directly. Therefore, the ratio obtained from the UrQMD reg configuration is used as the model counterpart of $\left[h^{*}/h\right]_{\mathrm{chemical}}$, while the ratio obtained from the UrQMD reg+res configuration is used as the model counterpart of $\left[h^{*}/h\right]_{\mathrm{kinetic}}$. Accordingly, the extracted $\tau_{\mathrm{had}}$ should be regarded as a model-dependent effective quantity defined from the difference between the regeneration-included and reconstructible resonance yields. It is therefore intended primarily for relative comparisons among particle species and collision systems.

As shown in Fig.~\ref{fig:fig_tau}, the estimated $\tau_{\mathrm{had}}$ generally increases with increasing multiplicity and system size for most resonance species. This behavior is consistent with the expectation that the hadronic phase survives longer in larger collision systems. Resonances with similar intrinsic lifetimes tend to yield comparable values of $\tau_{\mathrm{had}}$, whereas shorter-lived resonances systematically yield shorter estimated hadronic-phase lifetimes. This lifetime ordering is observed consistently among the resonance species considered in this study.
The observed ordering is also qualitatively consistent with the hadronic-phase lifetime extracted by the ALICE Collaboration using $\rho(770)^0$, $\mathrm{K}^{*}(892)^0$, and $\Lambda(1520)$~\cite{ALICE:2022wpn}. 

A comparison among collision systems reveals additional structure. In O--O and Pb--Pb collisions, $\tau_{\mathrm{had}}$ increases approximately linearly with $\langle dN_{\mathrm{ch}}/d\eta\rangle^{1/3}$. In contrast, in pp and p--O collisions, $\tau_{\mathrm{had}}$ increases at low multiplicities but tends to level off toward higher multiplicities. Although this feature is only weakly visible in the resonance-to-ground-state particle ratios themselves, it becomes more apparent after converting the ratios into $\tau_{\mathrm{had}}$. This behavior is also consistent with the $\langle p_{\mathrm{T}} \rangle$ trends discussed in the previous section, where pp and p--O collisions showed similar behavior, whereas O--O and Pb--Pb collisions formed a separate group.

This system dependence may arise from differences in the mechanisms that produce the same final-state multiplicity across different collision systems. In O–O and Pb–Pb collisions, increasing multiplicity is closely related to a larger overlap geometry and reaction volume. In contrast, in pp and p--O collisions, the collision geometry is more limited, and variations in energy density or particle-production density may play a more important role in determining the final-state multiplicity. Consequently, the effective size and lifetime of the hadronic phase in pp and p--O collisions may no longer increase monotonically with multiplicity. In the present calculations, the estimated $\tau_{\mathrm{had}}$ in these systems tends to level off at around 1 fm/$c$. It should be emphasized, however, that this interpretation is model dependent and requires further quantitative investigation.

The present lifetime extraction also has important limitations. The relation used above implicitly assumes that the reduction of the resonance-to-ground-state particle ratio is governed solely by rescattering. However, as demonstrated throughout this work, the resonance-to-ground-state particle ratio is also affected by regeneration and by the multiplicity dependence of the primary production yields. Therefore, for resonance species with sizable regeneration contributions, the observed modification of the ratio cannot be interpreted simply as a consequence of daughter-particle rescattering. The extracted $\tau_{\mathrm{had}}$ should therefore be regarded as an effective model observable rather than an absolute measurement of the hadronic-phase lifetime.

These results demonstrate that estimating the hadronic-phase lifetime from resonance-to-ground-state particle ratios requires accounting for both rescattering-induced signal suppression and regeneration-driven yield enhancement. A simultaneous treatment of these two competing mechanisms will be essential for future studies aiming to extract the hadronic-phase lifetime from resonance measurements.

\section{Summary}

\label{sec:Sum}
In this work, resonance production was systematically investigated using the EPOS4 event generator with and without the UrQMD hadronic cascade. A wide range of mesonic and baryonic resonances with different lifetimes and quark contents was studied in pp collisions at $\sqrt{s}=13.6$ TeV, p--O collisions at $\sqrt{s_{\mathrm{NN}}}=9.62$ TeV, and O--O and Pb--Pb collisions at $\sqrt{s_{\mathrm{NN}}}=5.36$ TeV. By comparing the UrQMD OFF, UrQMD reg, and UrQMD reg+res configurations, the effects of hadronic interactions on $p_{\mathrm{T}}$-integrated yields, resonance-to-ground-state particle ratios, and $\langle p_{\mathrm{T}} \rangle$ were investigated.

The integrated yields and resonance-to-ground-state particle ratios show that the final observable resonance yields are determined by the competition between regeneration and rescattering. The relative importance of these two processes varies significantly among different resonance species. While rescattering leads to substantial signal suppression for short-lived resonances, regeneration can be comparable to, or even dominate, rescattering for several baryonic resonances. These results demonstrate that hadronic-phase effects should not be interpreted simply as resonance suppression, and that regeneration must be considered as an essential component of resonance production in the hadronic phase.

The analysis of $\langle p_{\mathrm{T}} \rangle$, together with the $p_{\mathrm{T}}$-differential UrQMD ON/OFF ratios, provides further insight into the origin of these modifications. Regeneration predominantly enhances resonance production in the low-$p_{\mathrm{T}}$ region, thereby shifting the spectra toward lower transverse momentum and reducing $\langle p_{\mathrm{T}} \rangle$. In contrast, daughter-particle rescattering preferentially removes reconstructible resonance signals at low $p_{\mathrm{T}}$, thereby increasing the measured $\langle p_{\mathrm{T}} \rangle$. Rescattering exhibits a relatively clear dependence on the resonance lifetime, whereas both the magnitude and transverse-momentum dependence of regeneration show a much stronger particle-species dependence.

By extending the study to p--O and O--O collisions, a continuous system-size scan connecting pp and Pb–Pb collisions was performed. The integrated yields and resonance-to-ground-state particle ratios exhibit an overall multiplicity-driven evolution across the full system-size range, with p--O and O--O collisions naturally bridging the gap between pp and Pb--Pb collisions. In contrast, $\langle p_{\mathrm{T}} \rangle$ retains a stronger dependence on the collision system, suggesting that transverse-momentum distributions are sensitive not only to the final-state multiplicity but also to the underlying collision geometry, energy density, and initial-state dynamics.

The hadronic-phase lifetime was further estimated from the resonance-to-ground-state particle ratios. The extracted lifetime generally increases with system size, and resonances with similar intrinsic lifetimes yield similar values of the estimated hadronic-phase lifetime. At the same time, the present study shows that resonance-ratio-based lifetime estimates are affected not only by rescattering but also by regeneration and the multiplicity dependence of primary resonance production. Therefore, regeneration must be explicitly accounted for when interpreting hadronic-phase lifetime measurements.

A key outcome of this work is the separate investigation of regeneration and rescattering within a common model framework, which allows their individual roles to be identified. Rescattering shows a relatively clear dependence on the resonance lifetime and becomes more important in larger and higher-multiplicity systems. Regeneration, on the other hand, is strongly dependent on the particle species and can significantly modify both the yield and the transverse-momentum distribution of resonances. Consequently, hadronic-phase effects cannot be understood solely in terms of resonance lifetime. Instead, the balance between regeneration and rescattering, together with species-dependent hadronic interaction channels, must be considered simultaneously. These findings provide new insight into the interpretation of resonance observables and establish an important framework for future studies of hadronic-phase dynamics and lifetime extraction in small- and intermediate-collision systems.



\section{Acknowledgments}
H. Lim is supported by the National Research Foundation of Korea (NRF) grant funded by the Korean government (MSIT) under Contract No. RS-2025-25436721 and 2022 BK21 FOUR Program of Pusan National University.
H. Lim and S. Lim are supported by the National Research Foundation of Korea (NRF) grant funded by the Korean government (MSIT) under Project No. RS-2025-00554431 and RS-2008-NR007226. We also acknowledge technical support from KIAF administrators at KISTI.

\clearpage

\bibliography{main}

\end{document}